\documentclass[article]{llncs}
\usepackage{graphicx}
\usepackage{amsmath,amsthm,amssymb}

\usepackage{xcolor}

\usepackage{wrapfig}

\usepackage{multirow}
\usepackage{verbatim} 
\usepackage{booktabs}
\usepackage{graphicx}

\usepackage{wrapfig}

\usepackage{algorithmic,algorithm}

\usepackage[breaklinks=true,hidelinks]{hyperref} 

\usepackage{svg}
\newcommand{\orcid}[1]{\href{https://orcid.org/#1}{\includegraphics[width=10pt]{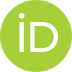}}}

\begin{document}
\title{Comparison of HPC Architectures for Computing All-Pairs Shortest Paths. \\Intel Xeon Phi KNL vs NVIDIA Pascal}
%
%
\author{Manuel Costanzo\inst{1}\orcid{0000-0002-6937-3943}
\and Enzo Rucci\inst{1}\orcid{0000-0001-6736-7358}\thanks{Corresponding author.} \and
Ulises Costi\inst{2}  \and
Franco Chichizola\inst{1}\orcid{0000-0001-8857-6343} \and
Marcelo Naiouf\inst{1}\orcid{0000-0001-9127-3212}}
%
%
\institute{III-LIDI, Facultad de Informática, UNLP – CIC. \\La Plata (1900), Bs As, Argentina \\
\email{\{mcostanzo,erucci,francoch,mnaiouf\}@lidi.info.unlp.edu.ar} \\
 \and
Facultad de Informática, UNLP. \\
La Plata (1900), Bs As, Argentina\\
\email{ulises\_92\_lp@hotmail.com}}
\maketitle              
\begin{abstract}
Today, one of the main challenges for high-performance computing systems is to improve their performance by keeping energy consumption at acceptable levels. In this context, a consolidated strategy consists of using accelerators such as GPUs or many-core Intel Xeon Phi processors. In this work, devices of the NVIDIA Pascal and Intel Xeon Phi Knights Landing architectures are described and compared. Selecting the Floyd-Warshall algorithm as a representative case of graph and memory-bound applications, optimized implementations were developed to analyze and compare performance and energy efficiency on both devices. As it was expected, Xeon Phi showed superior  when considering double-precision data. However, contrary to what was considered in our preliminary analysis, it was found that the performance and energy efficiency of both devices were comparable using single-precision datatype.

\end{abstract}

\noindent
\texttt{The final authenticated version is available online at {\url{https://doi.org/10.1007/978-3-030-75836-3\_3}}
}

\clearpage

\section{Introduction}
\label{sec:intro}


In the last decade, the quest to improve the energy efficiency of high-performance computing (HPC) systems has fueled the trend toward heterogeneous computing and massively parallel architectures~\cite{Giefers2016}. 
Heterogeneous systems combine CPUs with accelerators (such as NVIDIA and AMD GPUs or Intel's Xeon Phi many-core co-processors), delegating code sections with high computational demand to them. According to the Green500~\footnote{\url{https://www.top500.org/green500/}} ranking, the November 2010 edition featured 17 systems that integrated accelerators. However, 10 years later, this number has increased to 145, evidencing the great popularity of this strategy.

Today, GPUs can be considered the dominant accelerator class due to their high computing power and energy efficiency, in addition to their low cost. In the opposite sense, one of their weaknesses is the need to learn a specific language in order to make the most of them, such as CUDA and OpenCL. Among the manufacturing companies, NVIDIA stands out as the largest provider for the high-performance segment.

On the other hand, Intel introduced the second generation of its Xeon Phi processors in 2016, codenamed Knights Landing (KNL). Unlike its predecessor, Knights Corner (KNC), KNL can operate as a standalone processor. Among its main features, the large number of cores with hyper-threading support, the incorporation of AVX-512's 512-bit vector instructions, and the integration of a high-bandwidth memory (HBM), among others~\cite{KNLbook}, can be mentioned. Generations aside, the outstanding feature in this family is that it offers support for x86 architectures, which allows programmers to use traditional models in HPC, such as OpenMP and MPI.

Because each accelerator has its advantages and disadvantages for certain classes of problems~\cite{vestias2014,Deng2015,SWIFOLD}, selecting the best option for a given application is key when searching for maximum performance. To provide some guidelines for such selection, this article presents a comparative analysis between two different HPC architectures (Intel Xeon Phi KNL vs. NVIDIA Pascal). As a case study, the Floyd-Warshall (FW) algorithm was selected for computing the all-pairs shortest paths in a graph, as a representative case of graph applications that are memory-bound~\cite{RucciCACIC2017}. We hope that development teams will find this analysis useful when choosing the most suitable architecture for their applications.

The remaining sections of this article are organized as follows: in Section~\ref{sec:back}, the general background and other works that are related to this research are presented. Next, in Section~\ref{sec:imps}, the implementations used are described, and in Section~\ref{sec:results}, the experimental work carried out is detailed and the results obtained are analyzed. Finally, in Section~\ref{sec:conc}, our conclusions and possible lines of future work are presented.

\section{Background and Related Works}
\label{sec:back}

First, the Intel Xeon Phi KNL and NVIDIA Pascal architectures are briefly described and compared. Then, the FW algorithm for all-pair shortest paths computation in a graph is described. Finally, some works related to this article are detailed.

\subsection{Intel Xeon Phi KNL vs NVIDIA Pascal}
\label{subsec:arch}


\subsubsection{Intel Xeon Phi KNL.}

Unlike a GPU or a co-processor such as KNC, KNL is a standalone processor, capable of booting operating systems and directly accessing DDR memory. The scalable unit of replication of the KNL architecture is the \textit{tile}. Each tile houses 2 cores, a L2 cache shared between both cores, and a portion of the distributed directory. The cores of a tile implement \textit{simultaneous multi-threading} (4 hw threads per core) and out-of-order execution in its pipeline, in addition to having 2 vector units that support AVX-512's new 512-bit instructions~\cite{KNLbook}.

A KNL chip has between 32 and 36 active tiles (between 64 and 72 cores), depending on each specific model. The tiles are interconnected by a 2D mesh, with cache coherence based on distributed directory and MESIF protocol. This mesh can be configured in five different execution modes (cluster modes). Based on the execution mode selected, the distributed directory will be split among the tiles in the chip, which will have an impact on latency and bandwidth in memory access.

KNL comes with a 16GB HBM called MCDRAM, which is built into the same processor package. This memory can be configured in three different modes. In \textit{flat} mode, the address space is mapped between the two memories (MCDRAM and DDR), so it is the programmer who has the responsibility of defining how to use them. On the other hand, \textit{cache} mode leaves the system in charge of managing the MCDRAM as a DDR cache. Finally, the \textit{hybrid} mode assigns part of the MCDRAM as flat, and part as cache~\cite{knl_best_practice_guide}.

\subsubsection{NVIDIA Pascal.}






Pascal is the penultimate micro-architecture introduced by NVIDIA for the high-performance segment, successor to Maxwell. In this segment, a GPU of this family can have up to 3840 CUDA cores distributed among (at most) 60 Streaming Multiprocessors (SM). Compared to its predecessor, Pascal doubles the number of registers per core and increases the available size of shared memory.

This micro-architecture features several significant improvements over Maxwell. Among them, we can highlight the inclusion of an HBM of up to 16GiB in some of its chips, which allows them to reach a bandwidth of up to 720 GB/s. It also replaces the traditional PCIe bus used for CPU-GPU communication by a high-speed bus (NVLink), which significantly improves communication speed. Finally, the provision of a \textit{unified memory}, consisting of a virtual address space between the CPU and GPU memories, aimed at simplifying programming~\cite{PascalWP}.

As regards peak performance, some Pascal chips can achieve double-precision (DP) throughput rates that are half of the single-precision (SP) ones. On the other hand, they can double SP performance if they apply half-precision computing~\cite{Foley2017Pascal}.

\begin{table}[t]
\centering
\caption{Intel Xeon Phi KNL 7230 vs NVIDIA Titan X}
\label{tab:specs}
\resizebox{\textwidth}{!}{%
\begin{tabular}{p{3.8cm}  p{4.3cm}  p{4.3cm}}
\toprule
\textbf{\textbf{Device}} & \textbf{Intel Xeon Phi KNL} & \textbf{NVIDIA Titan X} \\ \midrule
\textbf{Chip} & 7230 & GP102 \\
\textbf{Clock Frequency} & 1.3-1.5 GHz & 1.42-1.53 GHz \\
\textbf{Cores} & 64 (256 hw threads) & 28 SMs (3584 CUDA cores) \\
\textbf{Cache} & 1 MB L2 & 3 MB L2 \\
\textbf{SIMD} & 512-bit & - \\
\textbf{HBM} & 16GB MCDRAM (450 GB/s) & - \\
\textbf{RAM Memory} & 192GB DDR4 (115.2 GB/s) & 12GB GDDR5X (480.4 GB/s) \\
\textbf{Bus} & - & PCI-Express 3.0 x16 \\
\textbf{Peak Theoretical performance SP (DP)} & 6 (3) TFLOPS & 10.97 (0.342) TFLOPS \\
\textbf{TDP} & 215W & 250W \\
\textbf{TFLOPS/W (SP/DP)} & 0.028 / 0.014 & 0.051 /  0.001 \\
\textbf{Launch Date} & June 2016 & August 2016 \\
\bottomrule
\end{tabular}%
}
\end{table}

 \subsubsection{Brief comparison.}
 
Table~\ref{tab:specs} presents a comparison of these architectures and, in particular, considers the models used in the experimental work. From the point of view of the theoretical peak performance in SP, the Titan X is far superior to the KNL 7230 (10.97 TFLOPS vs. 6 TFLOPS). However, due to the weak support of the former for DP, it is the KNL in this case that has a remarkable superiority (3 TFLOPS vs. 0.342 TFLOPS).
 
 As regards main memory, the KNL has an HBM that puts it almost on par with Titan X in terms of bandwidth, since KNL has DDR4 technology while Titan X has GDDR5X. However, since it is a co-processor, the Titan X has a much smaller memory size than the KNL, which is a \textit{host} unto itself.
 
Finally, when considering the (theoretical) energy efficiency, even though the Titan X has a higher thermal design power (TDP), its TFLOPS/W ratio in SP almost doubles that of KNL due to its higher theoretical performance peak. On the contrary, the poor performance of the Titan X for DP results in KNL being vastly superior in this case.
Despite the fact that some years have passed since their launch, both architectures remain relevant, as shown by the latest edition of the Top500~\footnote{Top500~\url{www.top500.org}} ranking, where 17 systems are equipped with Xeon Phi KNL and an additional, 30 with GPUs from the Pascal family.

\subsection{All-Pair Shortest Paths Computation in a Graph}

\subsubsection{FW Algorithm.}


The pseudocode of FW is shown in Algorithm~\ref{alg:algfw}. Given a graph \textit{G} of \textit{N} vertexes, FW receives as input a dense \textit{N}$\times$\textit{N} matrix $D$ that contains the distances between all pairs of vertexes from \textit{G}, where $D_{i,j}$ represents the distance from node \textit{i} to node \textit{j}~\footnote{If there is no path between nodes \textit{i} and \textit{j}, their distance is considered to be infinite (usually represented as the largest positive value)}. FW computes \textit{N} iterations, evaluating in the \textit{k}-th iteration all possible paths between vertexes \textit{i} and \textit{j} that have \textit{k} as the intermediate vertex. As a result, it produces an updated matrix \textit{D}, where $D_{i,j}$ now contains the shortest distance between nodes \textit{i} and \textit{j} up to that step. Also, FW builds an additional matrix \textit{P} that records the paths associated with the shortest distances.


\begin{algorithm}[t]
\label{alg:algfw}
\caption{Pseudocode of the FW algorithm}
\begin{algorithmic}
\FOR{$k \gets 0$ to $N-1$}
    \FOR{$i \gets 0$ to $N-1$}
        \FOR{$j \gets 0$ to $N-1$}
            \IF {$D_{i,j}$ $\geq$ $D_{i,k}$ + $D_{k, j}$} 
                \STATE $D_{i,j}$ $\gets$ $D_{i,k}$ + $D_{k, j}$
                \STATE $P_{i,j}$ $\gets$ $k$
            \ENDIF
        \ENDFOR
    \ENDFOR
\ENDFOR
\end{algorithmic}
\addtocounter{algorithm}{-1}
\end{algorithm}

\subsubsection{Blocked FW Algorithm.}

At first glance, the nested triple loop structure of this algorithm is similar to that of dense matrix multiplication (MM). However, since read and write operations are performed on the same matrix, the three loops cannot be freely exchanged, as is the case with MM. Despite this, the FW algorithm can be computed by blocks under certain conditions~\cite{floyd_blocked2}.

The blocked FW algorithm (BFW) divides matrix \textit{D} into blocks of size $TB\times TB$, totaling $(N/TB)^2$ blocks. Computation is organized in $R=N/TB$ rounds, where each round consists of 4 phases ordered according to the data dependencies between the blocks:

\begin{enumerate}
    \item Phase 1: Update the $D^{k,k}$ block because it only depends on itself.
    \item Phase 2: Update the blocks in row \textit{k} of blocks ($D^{k,*}$) because each of these depends on itself and on $D^{k,k}$. 
    \item Phase 3: Update the blocks in column \textit{k} of blocks ($D^{*,k}$) because each of these depends on itself and on $D^{k,k}$.  
    \item Phase 4: Update the remaining $D^{i,j}$ blocks of the matrix because each of these depends on blocks $D^{i,k}$ and $D^{k,j}$  on its row and column of blocks, respectively.
\end{enumerate}

\begin{figure}[b]
    \centering
    \includegraphics[width=1\textwidth]{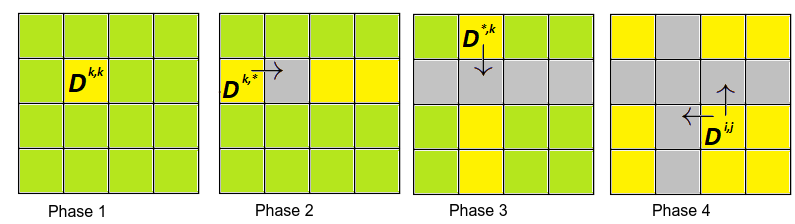}
    \caption{BFW computation phases and block dependencies}
    \label{fig:phases}
\end{figure}

Figure~\ref{fig:phases} shows each of the computation phases and the dependencies between blocks. The yellow squares represent blocks that are being computed, gray squares are those that have already been processed, and green squares are the ones that have not been computed yet. Last, arrows show the dependencies between blocks for each phase.

\subsection{Related Works}

The comparison of HPC architectures is a topic widely studied by the scientific community. In particular, there are several articles involving comparative studies between Xeon Phi KNL and NVIDIA Pascal in the field of large-scale economic modeling~\cite{Scheidegger2018} linear algebra~\cite{DEVECI2018}, computational fluid dynamics~\cite{Robertsen2019}, and automatic deep learning~\cite{gputitanx}, among others. However, as far as we know, this is the first to consider graph algorithms, in particular the FW algorithm for all-pair shortest paths.

The contributions of this work can be seen as an extension of a previous work of the authors~\cite{CostanzoCACIC2020}, where the comparison of HPC architectures just considered performance and theoretical energy efficiency. By incorporating actual power consumption and programming cost to the comparison, this work is able to offer a more comprehensive analysis of the pro and cons of each architecture for graph applications.
\section{FW Optimization}
\label{sec:imps}

This section describes the Xeon Phi implementation followed by the GPU one.

\subsection{Implementation on Xeon Phi KNL 7230}

The implementation used considers the following optimizations:

\begin{itemize}
    \item \textit{Data locality}. By computing with BFW, it is not only possible to exploit data locality, but it is also possible to increase the parallelism available in the application.
    \item \textit{Parallelism at thread level}. Using OpenMP, a multi-threaded version is obtained. Both Phase 2 and Phase 3 blocks are distributed among the different threads by means of the \texttt{for} directive with \texttt{dynamic} scheduling. In the case of Phase 1, since it consists of a single block, the iterations within it are distributed among the threads.
    \item \textit{Parallelism at data level}. Using the OpenMP \texttt{simd} directive, the operations of the innermost loop are vectorized when computing each block, which allows taking advantage of the AVX-512 instructions.
    \item \textit{Loop unrolling}. By fully unrolling the innermost loop and loop \textit{i} only once.
    \item \textit{Branch prediction}. By including the built-in \texttt{\_\_builtin\_expect} compiler macro, \texttt{if} statement branchs can be better predicted.
    \item \textit{MCDRAM}. Since this is a bandwidth-limited application, using this special memory is greatly beneficial. Executions are done using the \texttt{numactl} command.
\end{itemize}

It should be noted that this implementation can be considered as an optimized version of~\cite{RucciCACIC2017}, since it also includes intra-block parallelization for Phase 1 and the improvement in branch predictions.

\subsection{Implementation on NVIDIA Titan X}

The implementation used considers the optimizations known for GPU-based FW solutions at the moment~\cite{floyd_katz_gpu,MultiStageFW}. Among those, the following can be mentioned:

\begin{itemize}

    \item \textit{Concurrency}. Three kernels have been developed that are invoked once for each BFW computation round. On round \textit{k}, the first kernel computes block $D^{k,k}$ (Phase 1) and is instantiated with a single grid made up of a single block. Next, the second kernel is invoked, which computes both Phase 2 and Phase 3 blocks ($D^{k,*}$, $D^{*,k}$), and is instantiated with a single grid of $2\times(R-1)$ blocks. Finally, the third kernel, which computes the remaining blocks corresponding to Phase 4 ($D^{i,j}$), is invoked. This kernel is instantiated with a single grid of $(R-1)^2$ blocks. In all cases, blocks are made up of $TB\times TB$ threads.

    \item \textit{Exploitation of memory hierarchy}. For BFW computation, the use of shared memory is not only convenient but also necessary, especially in Phases 1-3 since the threads read and write to the same blocks of the matrix due to their dependencies. In the case of Phase 4, it is possible to take advantage of the private memory for the write block ($D^{i,j}$), which improves access times even more. Finally, the main memory accesses were organized so that they are coalescent.

    \item \textit{Resource occupation}. In order to optimize this aspect, different thread block sizes were tried (using $TB=\{8,16,32\}$) to find the one that leads to the maximum possible number of active \textit{warps}.
    
    \item \textit{Loop unrolling}. By fully unrolling the loop that computes each thread.
\end{itemize}

\section{Experimental Results}
\label{sec:results}

In this section, the experimental setup and methodology are described. Next, the results found are presented and analyzed. Last, the limitations of this research are mentioned.

\subsection{Experimental Setup and Methodology}

The tests were carried out on two different platforms~\footnote{The characteristics of each platform were described at the end of Section~\ref{subsec:arch}}. On the one hand, an Intel Xeon Phi KNL 7230 server configured in \textit{all-to-all} cluster mode and with \textit{flat} memory (ICC v19.0.0.117). On the other, an Intel Core i7-7700 3.6 GHz and 16GB RAM, which integrates an NVIDIA Titan X GPU (CUDA v9.0).

For both platforms, the variation in the size of the distances matrix ($N=\{4096,8192,16384,32768,65536\}$) and data type (\textit{float}, \textit{double}) were considered. In the case of KNL, different values for both the number of OpenMP threads $T=\{64,128,192,256\}$ and $TB=\{16,32,64,128\}$ were tried to identify the optimal values of $T_{float}=128$, $T_{double}=64$ and $TB=64$. As regards the GPU, the best performances were obtained when using $TB_{float}=32$ and $TB_{double}=16$. Finally, to minimize variability, each specific test was repeated 15 times and the average values were calculated.

Since this paper considers power consumption as well as performance, the measurement environment used on each platform is described as follows:
\begin{itemize}
	\item \emph{Intel Xeon Phi KNL}. Intel has developed the Intel PCM~\footnote{Intel Performance Counter Monitor: \url{http://www.intel.com/software/pcm}\ } (Performance Counter Monitor) to take power measurements on both Intel Xeon and Xeon Phi processors. With Intel PCM interface, any programmer can perform an analysis of CPU resource consumption by means of hardware counters.	
	\item \emph{NVIDIA Titan X}. In modern NVIDIA GPUs, the NVIDIA System Management Interface utility (\emph{nvidia-smi}~\footnote{NVIDIA System Management Interface: \url{https://developer.nvidia.com/nvidia-system-management-interface}\ }) can be used to query power consumption at runtime. This tool is based on the NVIDIA Management Library (NVML) and  is intended to help in the management and monitorization of NVIDIA GPU devices. 		
\end{itemize}


\subsection{Performance Results}

The GFLOPS metric is used for performance evaluation:

\begin{equation}
GFLOPS=\frac{2 \times N^3}{t \times 10^9 }
\end{equation}

 where \textit{N} is the size of the distances matrix, \textit{t} is execution time (in seconds), and factor 2 represents the number of floating-point operations required by each iteration of the innermost loop.

\begin{figure}[t]
\centering
\includegraphics[width=1\columnwidth]{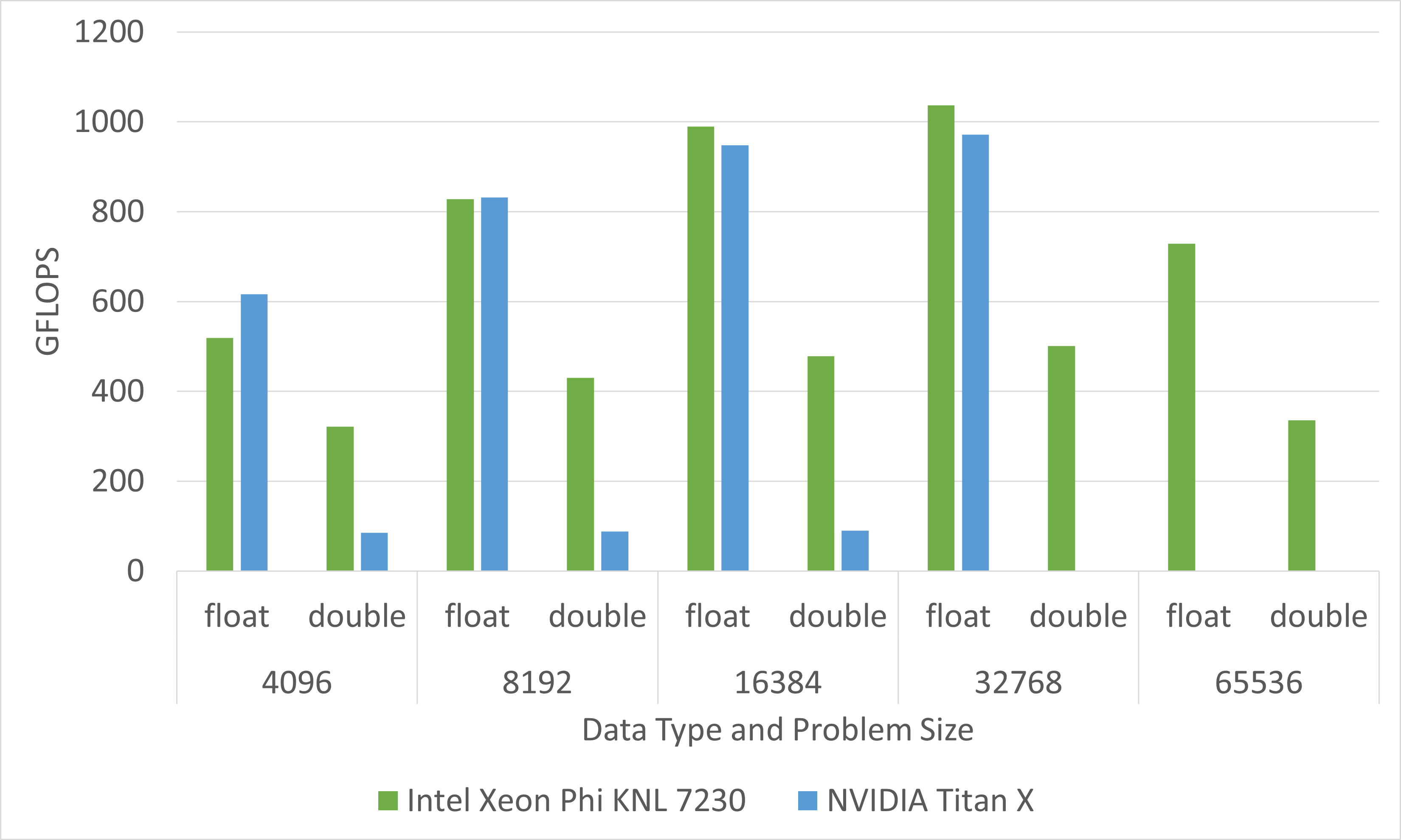}
\caption{Performance obtained with different data types and distances matrix sizes with Intel Xeon Phi KNL 7230 and NVIDIA Titan X}
\label{fig:flops}
\end{figure}

Figure~\ref{fig:flops} shows the performance obtained by each implementation with different values for both matrix and data type used. In SP (\textit{float}), it can be seen that the GPU achieves a better performance with smaller matrix sizes, being approximately 19\% higher when $N=4096$. In these cases, GPU is better suited than KNL, which requires higher workloads to reach its maximum use. This is reflected in the graph, since, as the size of the distances matrix increases, KNL achieves the best performance, reaching an additional 7\% difference. It should be noted that with $N=65536$, KNL experiences a performance loss of approximately 30\%, due to the fact that, with this value, the available size of the MCDRAM is exceeded and partial use of DDR4 is required, which has a much lower bandwidth. Even so, it is more convenient than using GPU, which cannot compute cases with $N>32768$ because the available main memory space is exceeded~\footnote{Naturally, it is also possible to develop an implementation that processes the matrix in parts and does not have this memory limitation. However, the need to run I/O operations for each round would significantly degrade performance.}. 

As regards DP (\textit{double}), the results obtained are as expected due to the weak support of Titan X for this class of operations. While GPU obtains an almost constant performance of $\sim$90 GFLOPS, KNL improves its performance as $N$ increases and eventually surpasses the size of the MCDRAM. In particular, KNL gets about half the FLOPS of SP but improves to 4.3$\times$ those of Titan X. Lastly, the memory limit issue in GPU is also worse due to the fact that the \textit{double} type requires more space (cases with $N>16384$ could not be processed).

\subsection{Power and Energy Efficiency Results}

As mentioned in Section~\ref{sec:intro}, application performance is not the only point of interest to be considered, energy efficiency also matters. Table~\ref{tab:perf-power-comp} lists energy efficiency ratios taking into account the GFLOPS peaks reached and the TDP of the platforms used. As it can be seen, KNL is superior in both cases. In particular, KNL gets a GFLOPS/Watt ratio that is 1.2$\times$ better than that of Titan X in SP. However, this factor increases up to 5.5$\times$ in DP, due to the weak support of this GPU for these operations. It should be noted that for this analysis, \textit{host} consumption by GPU was not taken into account, so the differences could be even greater.


TDP can be useful for qualitative comparison purposes and sometimes is the only valid metric when power measuring is not possible. However, actual power readings can differ from TDP due to a variety of reasons (power saving techniques available in modern processors/accelerators, specific workload in the target application, among others)~\cite{Igual-NNMF-2015}. For this reason, this work also includes an empirical power-performance analysis between platforms.

Table~\ref{tab:perf-real-power-comp} shows a summary of the best performance and the corresponding average power consumption on the
different architectures under study.

These power measurements reinforce the idea that actual power readings can differ from the TDP of each platform. In the KNL architecture, a higher power consumption is observed in both SP and DP scenarios; however, the difference is larger in the DP case. Hence, DP not only affects execution time but also increases power consumption, as it was also observed in other recent studies~\cite{Hashemi2017,Sakamoto2020}.

In opposite sense to the KNL case, average power consumption is lower than the corresponding TDP in Titan X. Considering this issue, Titan X becomes slightly better than KNL when SP is used.  Nevertheless, it must be remarked that host power consumption in not included in the GPU estimation; so, KNL still remains as probably the best option from this perspective. Despite the fact that energy efficiency gets almost doubled in DP, this improvement is virtually useless due to the poor performance of the Titan X.

\begin{table}[tb]
\centering
\caption{Theoretical comparison of performance and power efficiency by platform}
\label{tab:perf-power-comp}
\resizebox{\textwidth}{!}{%
\begin{tabular}{ccccc}
\toprule
\multicolumn{1}{l}{\textbf{Precision}} & \textbf{Platform} & \multicolumn{1}{l}{\textbf{GFLOPS (peak)}} & \textbf{TDP (Watt)} & \multicolumn{1}{l}{\textbf{GFLOPS/Watt}} \\
\midrule
\multirow{2}{*}{\textbf{SP}} & \textit{Xeon Phi KNL 7230} & 1037 & 215 & 4.82 \\
 & \textit{NVIDIA Titan X} & 972 & 250 & 3.89 \\
\midrule
\multirow{2}{*}{\textbf{DP}} & \textit{Xeon Phi KNL 7230} & 501 & 215 & 2.33 \\
 & \textit{NVIDIA Titan X} & 90 & 250 & 0.36 \\
 \bottomrule
\end{tabular}%
}
\end{table}

\begin{table}[tb]
\centering
\caption{Comparison of performance and power efficiency by platform}
\label{tab:perf-real-power-comp}
\resizebox{\textwidth}{!}{%
\begin{tabular}{ccccc}
\toprule
\multicolumn{1}{l}{\textbf{Precision}} & \textbf{Platform} & \multicolumn{1}{l}{\textbf{GFLOPS (peak)}} & \textbf{Power (Watt)} & \multicolumn{1}{l}{\textbf{GFLOPS/Watt}} \\
\midrule
\multirow{2}{*}{\textbf{SP}} & \textit{Xeon Phi KNL 7230} & 1037 & 229.7 & 4.51 \\
 & \textit{NVIDIA Titan X} & 972 & 209.5 & 4.64 \\
\midrule
\multirow{2}{*}{\textbf{DP}} & \textit{Xeon Phi KNL 7230} & 501 & 261 & 1.92 \\
 & \textit{NVIDIA Titan X} & 90 & 144.8 & 0.62 \\
 \bottomrule
\end{tabular}%
}
\end{table}

\subsection{Programming Cost}

As well as general-purpose architectures, KNL supports widely extended parallel programming models in HPC (such as OpenMP or MPI). On the other hand, CUDA is the \textit{de facto} standard for GPU programming nowadays. This fact puts KNL in a favourable position with respect to GPUs, since code development and portability get simplified.

Beyond specific language learning, GPUs may require additional programming efforts. Even though it experienced a significant performance loss when matrix size exceeded that of the MCDRAM, KNL was able to process all graphs. On the other hand, Titan X was not able to process those that exceeded the size of its RAM memory. While it is possible to develop an implementation that does, it would also have an associated performance loss and would require additional programming effort (not required in the case of KNL).

\subsection{Limitations}

Two possible limitations of this research can be mentioned:
\begin{itemize}
    \item  In 2019, Intel cancelled the KNL line~\cite{XeonPhiRequiem}. Even though, several features of there processors (like AVX-512 floating point unit, the mesh interconnect on the die, and the integration of high bandwidth stacked memory into the processor) have gone mainstream in the current Xeons~\cite{KNL_Legacy}. Thus, part of the results found in this research can be extrapolated to Xeon processors.
    
    \item This analysis focuses on two specific architecture models and that, as such, some of the results found could change if different models were used. For example, the NVIDIA GP100 (Pascal) GPU has a slightly lower peak SP performance than the Titan X (10.1 TFLOPS). However, its support for DP is vastly higher, being half SP (5 TFLOPS). In this sense, it is considered that the comparison is equally valuable to show trends and the distinctive characteristics of each architecture, even when there are different models that may present variations in their specifications.
\end{itemize}

\section{Conclusions and Future Work}
\label{sec:conc}

This work focuses on the comparison of Intel Xeon Phi KNL and NVIDIA Pascal architectures. Taking the FW algorithm for computing all-pairs shortest paths in a graph as a case study, optimized implementations were used to compare the achievable performance on each platform and thus be able to extract some general guidelines. Among those, the following can be mentioned:

\begin{itemize}

    \item Despite the fact that the preliminary analysis indicated that Titan X was far superior to KNL, the performances (SP) obtained for FW were comparable. While the GPU performed better for small graphs, as the size of the distances matrix increased, it was the KNL that performed better. This fact leads to KNL's need for large workloads in order to make the most of it.

    \item As regards energy efficiency, contrary to what was found in the preliminary analysis, no significant difference was observed in SP. This result contributes to the fact that device sustainable performance and energy efficiency vary depending on each particular problem and its corresponding software implementation.
    
    \item Beyond specific language learning, GPUs may require additional programming efforts due to their co-processor nature (they are not hosts per se) and limited memory sizes.

\end{itemize}




Future works include:

\begin{itemize}

    \item Extending the GPU implementation to support graphs larger than the main memory size.
    \item Including other models of the studied architectures  (especially Pascal GPUs).
\end{itemize}

The development of these activities would give greater robustness and representativeness to the study carried out.


\bigskip \noindent\textbf{Acknowledgments.} The authors are grateful for the support of NVIDIA through the donation of the Titan X GPU used in this research.

%
%
%
\bibliographystyle{splncs04}
\bibliography{references}
\end{document}